# Calorimetric Measurement of the Surface Energy of Enstatite, MgSiO$_3$

Megan A. Householder, Tamilarasan Subramani, Kristina Lilova, James R. Lyons, Rhonda M. Stroud, and Alexandra Navrotsky*

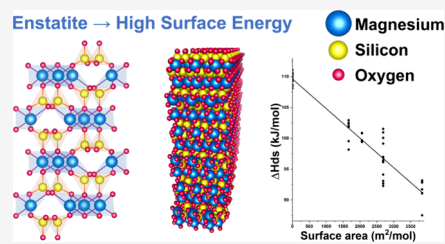

**ABSTRACT:** Surface thermodynamics of minerals influence their properties and occurrence in both terrestrial and planetary systems. Using high-temperature oxide melt solution calorimetry, we report the first direct measurement of the surface energy of enstatite, MgSiO$_3$. Enstatite nanoparticles of different sizes were synthesized using the sol−gel method, characterized with X-ray diffraction, thermal analysis, infrared spectroscopy, surface area measurements, and electron microscopy. The materials consist of crystallites with sizes of ∼10−20 nm, which are agglomerated into larger nanoparticles. Thus, both surface and interface terms contribute to the measured enthalpies. Analysis based on calorimetry and calculated surface and interface areas gives the surface enthalpy of enstatite as 4.79 ± 0.45 J m$^{−2}$. This value is comparable to that of forsterite (Mg$_2$SiO$_4$) and larger than those of many nonsilicate oxide materials. This large surface energy may present a barrier to the nucleation of enstatite in planetary atmospheres and other geochemical and planetary environments. The interfacial energy of enstatite appears to be close to zero. The transition enthalpy from bulk orthoenstatite to bulk clinoenstatite is 0.34 ± 0.93 kJ mol$^{−1}$, which is in agreement with earlier reports. The methodology developed here can be extended to other materials having complex structures and morphologies to separate surface and interfacial contributions to energetics.

## 1. INTRODUCTION

Surface energies are essential to understanding the primary and fundamental behaviors and interactions between chemical species at the surfaces of solids and liquids. Grain boundaries between particles and interfaces between crystallites have become important in ceramics and the semiconductor industry.[1−4] Surfaces in nanophase materials are the loci for most chemical reactivity, and surface energies determine surface reactivity, phase transformation, and grain growth.[5] These processes control important processes in nature such as growth of crystalline materials within rocky planets, nucleation and condensation of solids in planetary atmospheres, shock wave condensation from supernovae, and planetary formation through condensation within accretion disks around stars. The atmospheres of warm and hot exoplanets are likely to contain cloud condensates and photochemical hazes formed from different chemical species including enstatite.[6] The nucleation rate of such condensates depends strongly on their surface properties, with higher surface energies drastically hindering nucleation.[7] Although nucleation has been modeled extensively by the materials science, geochemistry, and astrophysics communities, the surface energies of many key compounds have not been constrained by experiments.

Enstatite is the magnesium endmember of the silicate pyroxene mineral series (MgSiO$_3$−FeSiO$_3$). Pyroxenes are rock-forming minerals with the crystal structure XYZ$_2$O$_6$.[8] Enstatite has three known distinct crystalline polymorphs: orthoenstatite (OEn), protoenstatite (PEn), and clinoenstatite (CEn). CEn has two forms, high-clinoenstatite (HCEn) and low-clinoenstatite (LCEn). The characteristics of these polymorphs have been studied extensively, but they are still far from being completely understood. OEn and PEn are both orthorhombic, with LCEn being monoclinic. OEn, space group $Pbca$, is stable at room temperature and ambient conditions but transforms to low-clinoenstatite upon grinding or under shear stresses, suggesting a martensitic transformation.[9] PEn, space group $Pbcn$, is reported to form from OEn and be the stable polymorph at ambient pressure from about 1000 to 1300 °C and possibly to the melting point at 1557 °C.[10] Upon cooling, PEn transforms to LCEn at 865 °C.[9] LCEn, space group $P2_1/c$, is regarded as metastable at room temperature and ambient pressure, and high-pressure HCEn, space group $C2/c$, is the stable phase above 6 GPa.[11]

Here, we report the surface energy of enstatite based on the oxide melt solution calorimetry of nanophase and bulk enstatite samples. Electron microscopy (EM) was used to determine the particle size to calculate the surface area (SA).

Table 1. Drop Solution Enthalpies of Enstatite in Lead Borate at 700 °C and Enthalpies of Formation from Oxides and Elements at 25 °C[a]

| compound | $\Delta H_{ds}$, kJ/mol | $\Delta H_{f,ox}$, kJ/mol | $\Delta H_{f,el}$, kJ/mol |
|---|---|---|---|
| bulk ortho MgSiO$_3$ | 109.93 ± 0.60 (8) | −34.05 ± 0.88 | −1546.35 ± 1.36 (this work), −1545.60 ± 1.50[23] |
| bulk low-clino MgSiO$_3$ | 109.59 ± 0.71 (8) | −33.71 ± 0.95 | −1546.01 ± 1.41 (this work), −1545.0 ± 1.50[23] |
| nano 1075 °C 48 h | 101.23 ± 1.18 (8) | −25.35 ± 1.34 | −1537.65 ± 1.70 |
| nano 1000 °C 24 h | 99.97 ± 0.56 (6) | −24.09 ± 0.85 | −1536.39 ± 1.35 |
| nano 900 °C 8 h | 97.61 ± 1.72 (13) | −21.73 ± 1.84 | −1534.03 ± 2.11 |
| nano 800 °C 2 h | 91.49 ± 1.47 (7) | −15.61 ± 1.61 | - 1527.91 ± 1.92 |

[a]The number of experiments is in parentheses; uncertainty is two standard deviations of the mean.

Phase percentages of the OEn and LCEn in several nanophase samples with different heating temperatures and calcination times were determined by Rietveld refinement of powder X-ray diffraction (PXRD) data. Because the nanophase samples were a mixture of ortho and clino polymorphs, we determined an average surface energy that should be applicable to both polymorphs.

## 2. MATERIALS AND METHODS

**2.1. Synthesis.** The starting materials for the synthesis of bulk and nanophase enstatite were tetraethyl orthosilicate (TEOS) (Sigma-Aldrich), magnesium nitrate hexahydrate (Acros Organics), and ammonium hydroxide (28%, VWR Chemicals). The magnesium nitrate hexahydrate (Mg(NO$_3$)$_2$ · 6H$_2$O) was dried in a muffle furnace at 100 °C for 12 h before use to remove any adsorbed water so the exact magnesium content could be determined to obtain the correct stoichiometry during synthesis. A similar decomposition reaction was also done on TEOS to determine its silicon content. The materials were then used as received, and all weighing was done using a semimicrobalance (Mettler Toledo) to ensure the stoichiometric Mg:Si ratio of 1:1 for the synthesis of MgSiO$_3$ without forsterite (Mg$_2$SiO$_4$) or MgO phases.

Synthesis used the sol−gel method.[12] Magnesium nitrate hexahydrate (Mg(NO$_3$)$_2$ · 6H$_2$O) was dissolved in a beaker with a stir bar in 50 mL of ethanol for 30 min, and TEOS was dissolved in ethanol in a separate beaker and stirred. The two solutions were combined and stirred for 1 h. Then, 50 mL of ammonium hydroxide (NH$_4$OH) was added, and the entire mixture was stirred for an additional hour. The resulting precursor mixture was dried on a hot plate for several days. When the mixture was dry, it was scraped from the beaker into a mortar and pestle and ground into a fine powder. To remove any remaining organics and water, the starting material was put in a platinum crucible in a temperature-controlled muffle furnace at 100 °C and heated at a rate of 60 °C/h to 600 °C with a dwell time of 2 h. This was the starting material for enstatite nanoparticles. To crystallize the starting material, 250 mg was heated to 800 °C for 2 h and cooled back to 100 °C. Three more nanophase samples with different particle sizes were prepared similarly but with different temperatures and dwell times for a total of four samples: 800 °C for 2 h, 900 °C for 8 h, 1000 °C for 24 h, and 1075 °C for 48 h.

Bulk LCEn was synthesized by taking the enstatite starting material and coarsening it into bulk particles by annealing in a platinum crucible at 1500 °C for 12 h in a Deltech MoSi$_2$ vertical furnace.

Bulk OEn was provided by Richard Hervig at Arizona State University.[13] It had been synthesized in a flux of MoO$_3$−Li$_2$O−V$_2$O$_5$ by Minoru Ozima of the University of Tokyo and provided to Navrotsky and Hervig by Eiji Ito of Okayama University. Impurities in it were <1%.[14] Because of the age of this material, it was characterized again, and we found that its structure had not changed in nearly 40 years.

**2.2. Characterization.** Thermogravimetry and differential scanning calorimetry (TG−DSC) were performed on a Setaram Labsys Evo instrument in argon with a heating rate of 2 °C/min to 1200 °C, followed by cooling at 20 °C/min.

PXRD was done on a Bruker D2 benchtop diffractometer equipped with Ni-filtered Cu $K_\alpha$ radiation ($\lambda$ = 1.542 Å) operating at an accelerating voltage of 20 kV and an emission current of 10 mA. The data were collected in the 2$\theta$ range of 10−80° with a 0.010 step size and a 2 s/step collection time. The PXRD patterns were analyzed using Rietveld refinement with GSAS-II software to determine the phase percentages of OEn and LCEn in the nanophase samples, phase purity of the bulk samples, and crystallite size based on peak broadening.[15] Fourier transform infrared spectroscopy (FTIR) was performed with attenuated reflectance spectroscopy (ATR) on a Bruker Vertex 70 spectrometer in the range of 100−6000 cm$^{-1}$.

The SA was measured by N$_2$ adsorption using a 10-point Brunauer−Emmett−Teller (BET) technique at −196 °C on the analysis port of a Micromeritics ASAP 2020 instrument calibrated with a previously characterized alumina (Al$_2$O$_3$) nanoparticle powder.[16] The powder samples were degassed at 100 °C in vacuum for 24 h before analysis. The average particle size was calculated from the SA using standard equations.[17]

Scanning transmission electron microscopy (STEM) was performed using a CM200-FEG high-resolution Phillips TEM/STEM at 200 keV. Transmission electron microscopy (TEM) was performed using a JEOL TEM/STEM ARM (Atomic Resolution Microscope) 200F at 80 keV and 4 pA. Scanning electron microscopy (SEM) was performed using a Thermo Scientific SEM/FIB Focused Ion Beam Helios 5 UX at 2 keV. For SEM, the samples were coated with Au/Pd (90:10) in a Denton Vacuum DESK II Cold Sputter/Etch Unit to prevent the sample from charging while taking secondary electron images. The SA was determined using the cross-line intersection method.[18,19] Over 400 particles from each sample were measured.

Further details of characterization methodologies and results are given in the Supporting Information (SI).

**2.3. Enthalpy of Formation Measurements.** High-temperature oxide melt drop solution calorimetry was performed at 700 °C in molten lead borate (2PbO·B$_2$O$_3$) solvent. A 5 mg pellet was dropped into the solvent for each measurement. The measured enthalpy is the sum of the heat of the solution and the heat content. The calorimeter was calibrated with 5 mg of corundum pellets. This methodology is standard in our laboratory and has been applied to many mineral phases and nanoparticles.[20−22] Drop solution enthalpies and thermodynamic cycles are given in Tables 1 and 2, respectively.

**Table 2. Thermodynamic Cycles Used to Calculate the Enthalpies of Formation from Oxides and from Elements[a]**

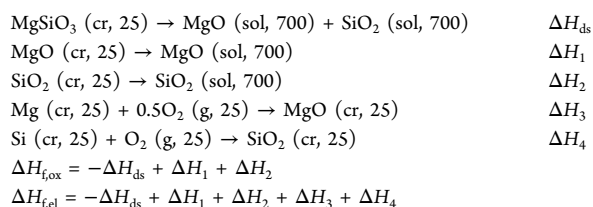

| | |
|---|---|
| $MgSiO_3$ (cr, 25) → MgO (sol, 700) + $SiO_2$ (sol, 700) | $\Delta H_{ds}$ |
| MgO (cr, 25) → MgO (sol, 700) | $\Delta H_1$ |
| $SiO_2$ (cr, 25) → $SiO_2$ (sol, 700) | $\Delta H_2$ |
| Mg (cr, 25) + 0.5$O_2$ (g, 25) → MgO (cr, 25) | $\Delta H_3$ |
| Si (cr, 25) + $O_2$ (g, 25) → $SiO_2$ (cr, 25) | $\Delta H_4$ |
| $\Delta H_{f,ox} = -\Delta H_{ds} + \Delta H_1 + \Delta H_2$ | |
| $\Delta H_{f,el} = -\Delta H_{ds} + \Delta H_1 + \Delta H_2 + \Delta H_3 + \Delta H_4$ | |

[a]Temperatures are in °C. The "cr" refers to crystal, "sol" refers to solution, and "g" refers to gas.

## 3. RESULTS AND DISCUSSION

**3.1. Synthesis.** TG−DSC of synthesized starting materials showed a broad exothermic peak from 660 to 845 °C with a maximum at 805 °C. Another sharp peak was observed from 845 to 880 °C with a maximum at 850 °C, implying a crystallization temperature range of 700 to 900 °C (Figure 1). We suspect that these two peaks correspond to LCEn and OEn crystallization, not necessarily in that order.

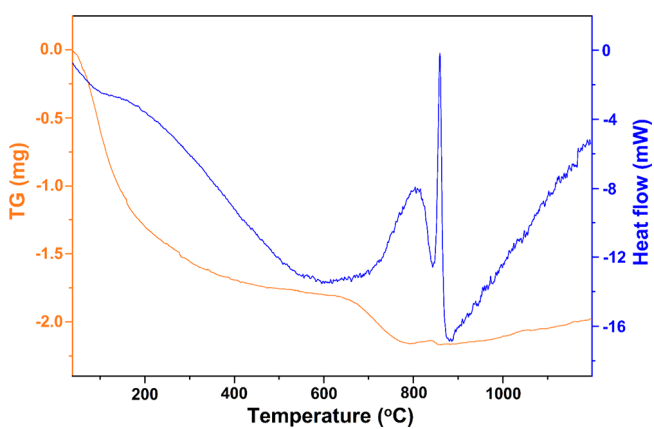

**Figure 1.** TG−DSC of the starting material. The conditions were stabilization at 25 °C for 30 min in argon gas, heating at 2 °C/min to 1200 °C with a dwell time of 1 h, and cooling at 20 °C/min to 25 °C. The mass loss (mg) is in orange and the heat flow (mW) is in blue. A broad exothermic peak was observed from 660 to 845 °C with a maximum at 805 °C. Another sharp peak was observed from 845 to 880 °C with a maximum at 850 °C, implying a crystallization temperature range of 700 to 900 °C.

The FTIR spectra showed that all nanocrystalline samples were free of water and organics. The expected peaks for enstatite were observed (see the Supporting Information), and peaks corresponding to other phases were absent.

XRD was performed on all samples to determine the phase purity, crystallite size, and phases present in the nanophase and bulk samples. All nanophase samples crystallized into nanoparticles containing both LCEn and OEn. The two bulk samples were pure LCEn and pure OEn.

Figure 2 shows the XRD patterns with increasing peak intensity, corresponding to increasing temperature and dwell times. Table 3 lists the nanophase sample preparation

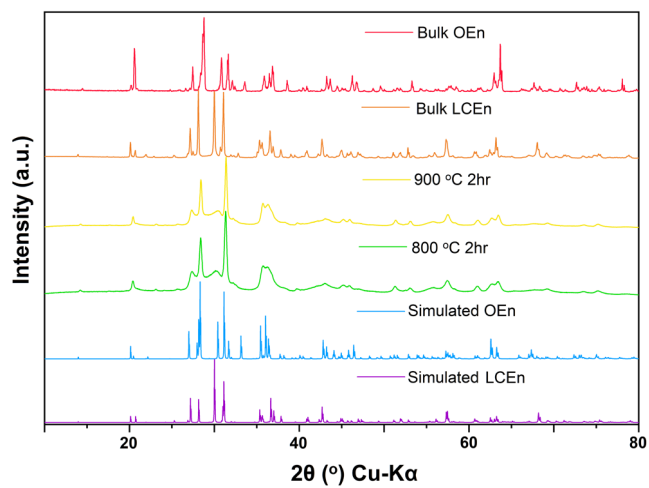

**Figure 2.** XRD patterns of the 800 and 900 °C synthesized nanophase and bulk enstatite samples and simulated patterns of OEn and LCEn.

conditions, OEn fractions, and corresponding LCEn and OEn crystallite sizes.

**3.2. Calorimetry.** The drop solution enthalpies and enthalpies of formation from oxides and elements of the enstatites are shown in Table 1. The drop solution enthalpy of the OEn is consistent with the solution enthalpy value from Hervig et al. (1985)[13] and previously determined heat content.[23] The thermodynamic cycles used to calculate the enthalpies of formation are shown in Table 2.

The measured standard formation enthalpy from the elements for bulk OEn is −1546.35 ± 1.36 kJ/mol, within an error of the tabulated value of −1545.6 ± 1.50 kJ/mol.[23] The measured formation enthalpy for bulk LCEn, −1546.01 ± 1.41 kJ/mol, is also within an error of that given in Robie and Hemingway (1995),[23] −1545.0 ± 1.5 kJ/mol.

The transition enthalpy of bulk OEn to bulk LCEn, calculated as the difference between the drop solution enthalpies of each phase, is 0.34 ± 0.93 kJ/mol. The transition enthalpy as calculated from previously published enthalpies of formation is 0.40 ± 2.10 kJ/mol,[23] within an error of the value in this work.

**3.3. Morphology and Surface and Interface Area.** Studies by XRD, EM, and BET strongly suggest that three different length scales are involved in describing the particle size and SA (see Table 3 and further discussion below). EM confirmed the presence of particle agglomerates (Figure 3a−c and Supporting Information). The average size of the agglomerates seen using EM was in general agreement with that calculated from the BET SA.

There are three length scales that increase in the following order: crystallite size determined by XRD, particle size determined by EM, and agglomerate size determined from BET and by EM. The crystallites identified by XRD have a coherence length of 10−20 nm and represent basic enstatite building blocks. These crystallites, rather than being individual particles in contact only with air, aggregate into nanoparticles containing both OEn and LCEn with dimensions in the 50−100 nm range, as seen by SEM and TEM. EM also showed that these nanoparticles formed agglomerates consistent with the BET SAs. The outside of such agglomerates adsorb the $N_2$ gas used for BET measurements, but apparently, the gas does not penetrate the grain boundaries between the crystallites forming the nanoparticles. Thus, the EM and BET studies provide the average nanoparticle size and external (particle to gas phase) area, which we call SA. It does not include the area between crystallites in contact with each other with different orientations and different interactions. For simplicity in further thermodynamic calculations and because we do not have quantitative data on types, numbers, and areas of different interfaces, we use one parameter to describe an average interfacial area (IA), which we call IA. The small crystallite size from XRD provides the area of surfaces plus interfaces, which we call total area, or TA. This TA is the sum of surface (crystallite−gas) and interface (crystallite−crystallite) areas: TA = SA + IA. Further details of these calculations are given in the Supporting Information. Since the TA is calculated from the XRD crystallite size and the SA from the EM measurements, one can calculate the IA as IA = TA − SA. These three different areas are shown in Table 3.

Table 3. XRD Crystallite Size, Fraction of OEn, TA from XRD, Average Particle Size from EM, SA from EM, Agglomerate Size and SA from BET, and IA of Nano-Enstatite Samples[a]

| temperature, time (°C, hr) | crystallite size (nm) (XRD) | fraction of OEn (XRD) | TA per phase (m$^2$/g) (XRD) | TA, (m$^2$/g) (XRD) | particle size (nm) (EM) | SA (m$^2$/g) (EM) | agglomerate size (nm) (BET) | SA (m$^2$/g) (BET) | IA (m$^2$/g) (XRD-EM) |
|---|---|---|---|---|---|---|---|---|---|
| 1075, 48 | LCEn: 16.8 | 0.772 | LCEn: 25.5 | 111.6 | 113.3 | 16.6 | | <1 | 95.0 |
|  | OEn: 16.8 |  | OEn: 86.2 |  |  |  |  |  |  |
| 1000, 24 | LCEn: 15.8 | 0.62 | LCEn: 45.2 | 107.3 | 91.4 | 20.5 | | <1 | 86.8 |
|  | OEn: 18.7 |  | OEn: 62.2 |  |  |  |  |  |  |
| 900, 8 | LCEn: 11.4 | 0.455 | LCEn: 91.3 | 128.5 | 70.0 | 26.8 | 457 | 4.10 | 102 |
|  | OEn: 22.9 |  | OEn: 37.3 |  |  |  |  |  |  |
| 800, 2 | LCEn: 10.9 | 0.419 | LCEn: 99.9 | 140.4 | 49.2 | 38.3 | 2400 | 0.78 | 102 |
|  | OEn: 19.4 |  | OEn: 40.5 |  |  |  |  |  |  |

[a]Uncertainties are reported in the Supporting Information.

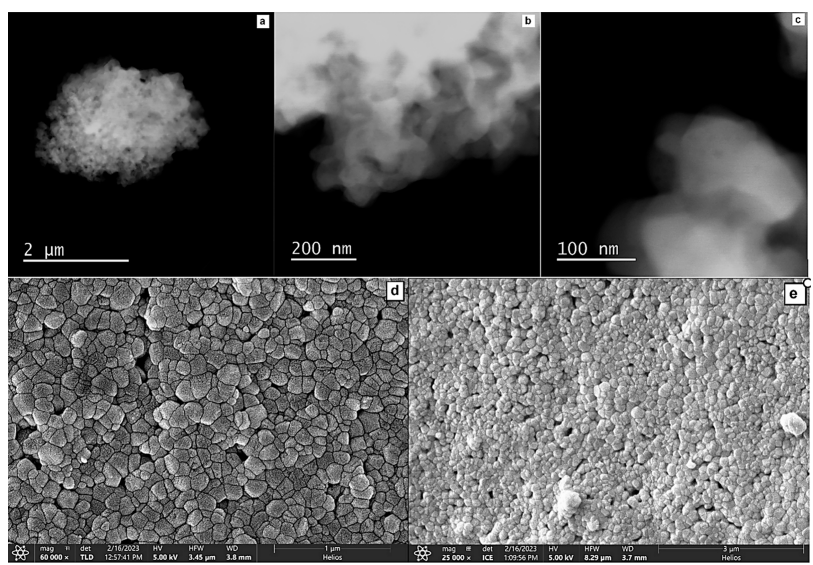

**Figure 3.** (a−c) HAADF STEM images taken at 200 keV and 4.0 pA of a 800 °C 2 h nanoparticle agglomerate, confirming BET data of an agglomerate with an exposed SA equivalent to ∼2.5 μm with (a) a scale bar of 2 μm, (b) a scale bar of 200 nm, and (c) a scale bar of 100 nm. (d, e) Secondary electron SEM image of the 1075 °C 48 h sample with (d) a scale bar of 1 μm and (e) a scale bar of 3 μm.

Once the area was found for each polymorph, it was multiplied by its phase fraction and then both added to obtain the TA (IA + SA). The XRD crystallite size, phase fraction of OEn and LCEn, and calculated TA from Rietveld refinement using GSAS-I software are shown in Table 3.

Rietveld refinement shows that the fraction of OEn increases with increasing temperature and time, and the crystallite size of LCEn increases with increasing temperature, while the crystallite size of OEn decreases. It appears that LCEn prefers lower temperatures and may grow and transform to OEn at higher temperatures.

It is clear that the 800 °C 2 h sample has an intricate pore system, and smaller particles than the other samples, but still has larger particle sizes than the XRD crystallite size. All of the nanophase samples have similar XRD crystallite sizes. These crystallites compose the larger particles, seen in the STEM, TEM, and SEM images (Figure 3 and Supporting Information).

Particle sizes contributing to the agglomerates were measured from EM images using the cross-line intersection method.[18,19] Over 400 particles from four different agglomerates from each sample were measured. Because the XRD crystallite size is ∼10−20 nm, but the particle sizes composing the agglomerates are larger (∼50−100 nm), we must consider the possibility that interfacial energy arising from crystallite−crystallite contact contributes to excess enthalpy in the calorimetry measurements; see the next section.

**3.4. Surface Energy Calculation.** The morphological complexity described above must be considered in the calculation of the surface energy. The measured excess enthalpy of the nanophases relative to a mixture of bulk LCEn and bulk OEn contains contributions from both free surfaces and interfaces. We calculated the IA and SA, as described above. We can approximate the total measured excess enthalpy, $\Delta H_{ex}$, as coming from these two sources. Then, the excess enthalpy is given by

$$\Delta H_{ex} = \Delta H_{ds}(\text{bulk}) - \Delta H_{ds}(\text{nano})$$
$$= (IA \times IE) + (SA \times SE) \quad (1)$$

where IE is the interface energy and SE is the surface energy.

Table 4. Pairwise Calculation of IE and SE[a]

| nano sample pair (°C) | 1075/1000 | 1075/900 | 1075/800 | 1000/900 | 1000/800 | 900/800 |
|---|---|---|---|---|---|---|
| SE (J/m$^2$) | 3.70 | 4.25 | 5.13 | 5.50 | 5.77 | 5.83 |
| IE (J/m$^2$) | 0.26 | 0.17 | 0.00 | −0.16 | −0.23 | −0.25 |

[a]Uncertainties are reported in the Supporting Information.

We then know the IA and SA for all four nanophase samples and can solve for the interfacial and surface energies. Attempts to fit all four samples simultaneously showed divergent results, so we used pairs of samples to obtain the energies, taking samples pairwise (Table 4). The interfacial energy was approximately zero (0.03 ± 0.39 J/m$^2$), and the surface energy was 3.70 to 5.83 J/m$^2$, with an average of 5.0 ± 0.71 J/m$^2$. The error is two standard deviations from the mean.

The calculations give a surface energy of 5.0 ± 0.71 J/m$^2$ and an interface energy of 0.03 ± 0.39 J/m$^2$. Thus, despite a significant interface area, the contribution of interfacial energy appears low because of its very small value calculated by the pairwise calculation. The physical reason for such a small interfacial energy is not clear, but it may suggest that the crystallites are strongly bonded to each other, minimizing an unfavorable interaction at crystallite–crystallite boundaries. Because of the emphasis on surface energy, TEM studies directly visualizing individual crystallites are beyond the scope of this work, although further efforts in this area would be desirable for interfacial energy studies.

Because of the very minor interface energy contribution, we then fit all the calorimetric data to the SAs measured by EM using a linear fit, which gave a surface energy of 4.79 ± 0.45 J/m$^2$ (see Figure 4).

The surface energy obtained refers to the mixtures of OEn and LCEn described above and should be considered an average value applicable to both polymorphs. Cumulative uncertainties in enthalpies, phase fractions, and SAs preclude meaningful calculation of the surface energy of each polymorph separately. However, the similarity in energetics and the structure of the bulk polymorphs suggests that both would have similar surface energies.

Table 5 compares the surface energies of several mineral phases. The surface energy of enstatite is close to that of forsterite (Mg$_2$SiO$_4$). It appears that silicate minerals have higher surface energies than simple oxides, especially spinels. However, data on additional silicate minerals are needed to confirm such trends.

Table 5. Comparison of Surface Energies for Several Oxides

| material | measured surface enthalpy for the anhydrous surface, J/m$^2$ |
|---|---|
| MgSiO$_3$ (enstatite) | 4.79 ± 0.45[a] |
| Mg$_2$SiO$_4$ (forsterite) | 4.41 ± 0.21[b] |
| α-Al$_2$O$_3$ | 2.64 ± 0.20[c] |
| TiO$_2$ (rutile) | 2.22 ± 0.07[d] |
| γ-Al$_2$O$_3$ (defect spinel) | 1.67 ± 0.10[c] |
| α-Fe$_2$O$_3$ | 1.90 ± 0.30[e] |
| γ-Fe$_2$O$_3$ (defect spinel) | 0.99 ± 0.25[f] |
| MgAl$_2$O$_4$ (spinel) | 1.80 ± 0.30[g] |

[a]This study. [b]Chen and Navrotsky 2010.[19] [c]McHale et al. 1997.[24] [d]Levchenko et al. 2006.[25] [e]Mazeina and Navrotsky 2007.[26] [f]Bomati-Miguel et al. 2008.[27] [g]McHale et al. 1998.[28]

**3.5. Implications beyond Earth.** Enstatite is a common mineral in many astrophysical as well as geochemical environments. In astrophysical environments, submicrometer to micrometer-sized Mg-rich pyroxenes are observed in interplanetary dust particles.[29] The potential for water adsorption into these particles and other Mg-rich refractory silicates has led astronomers to new theories of the origin of water on terrestrial planets.[30] Infrared observation of protoplanetary disks often shows evidence for silicate dust, enstatite and/or forsterite, at wavelengths near 10 μm.[31] Thermochemical equilibrium condensation models predict two zones of enstatite grain formation in the inner solar nebula that may correspond to the zones of formation for Mercury and the enstatite chondritic meteorites.[32]

Enstatite is a major component of the enstatite chondrites, which represent a reservoir of materials formed under reducing conditions in the early solar system. A study of enstatite chondrites, the only chondrites isotopically identical to Earth, has led to new models of Earth's chemical composition, further constraining models of Earth's core and its mantle.[33,34] These chondrites are important in studying the early history of the solar system, with age dates suggesting formation just a few million years after CAI (calcium–aluminum-rich inclusions) formation in meteorites. These CAIs are considered the first solids formed in the solar system.[35] Enstatite chondrites can also aid in the understanding of the composition of other rocky planets. Enstatite chondrites have also been considered possible analogues for Mercury, with its surface composition including some nearly FeO-free enstatite and FeO.[36] Clearly, enstatite was present in the solar nebula. However, the high surface energy for enstatite presented here suggests that it may not have condensed directly into the crystalline form.

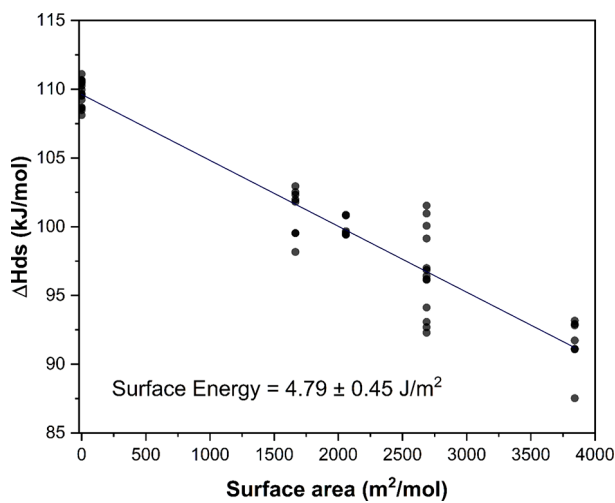

Figure 4. Surface energy of enstatite fitted as a function of the SA calculated from particle sizes using EM.

Outside our solar system, enstatite and other silicates are important in understanding exoplanetary atmospheres. The James Webb Space Telescope (JWST) has identified exoplanets with silicate clouds composed of enstatite, forsterite, and/or quartz.[37] The best fitting model of newly observed planetary-mass companion VHS 1256b has a cloud composition of enstatite, quartz, and high-pressure iron.[38] Silicates are also present in hot Jupiter and brown dwarf exoplanet aerosol clouds. Like the solar nebula, our results suggest that crystalline enstatite does not condense directly in such hot exoplanet atmospheres compared to condensates with lower surface energies. Surface energetics exponentially affect nucleation and condensation and thus are extremely important in modeling exoplanet atmospheres and interpreting exoplanet atmosphere data from the JWST. The need for experimentally measured surface energetics has never been more critical for the astrophysical community.

## 4. CONCLUSIONS

Despite particle agglomeration in enstatite nanoparticles leading to complexity in reconciling SA measurements by different techniques, careful analysis produced a consistent picture and enabled the calculation of the surface energy of enstatite to be $4.79 \pm 0.45$ J m$^{-2}$. This value is higher than those for many binary oxides but similar to that for forsterite. The high surface energies of enstatite and forsterite likely inhibit the direct condensation of crystalline nanoparticles of these two key silicates in a wide range of astrochemical environments. More measurements of the surface energies of rock-forming minerals, particularly silicates, are needed. The measured surface energies and trends with structure and composition that they define have strong cross-disciplinary implications for solid-state chemistry, materials science, geochemistry, planetary science, and astrophysics.

## ASSOCIATED CONTENT

### Supporting Information

The Supporting Information is available free of charge at https://pubs.acs.org/doi/10.1021/acs.jpcc.3c04211.

> Additional details about surface energy calculations, surface area calculations and equations, (Figures S1–S7) FTIR spectra, (Figure S8) XRD patterns of all samples, (Table S1) PXRD analysis results from GSAS software, (Table S2) uncertainties from Table 3, (Table S3) uncertainties from pairwise calculations, (Figure S9) weight percentage of LCEn vs OEn as a function of annealing temperature, and (Figures S10–S12) additional TEM/SEM images (PDF)

## AUTHOR INFORMATION

**Corresponding Author**

Alexandra Navrotsky − School of Earth and Space Exploration and Center for Materials of the Universe, Arizona State University, Tempe, Arizona 85287, United States; Center for Materials of the Universe and School of Molecular Sciences, Arizona State University, Tempe, Arizona 85281, United States; orcid.org/0000-0002-3260-0364; Email: Alexandra.Navrotsky@asu.edu

**Authors**

Megan A. Householder − School of Earth and Space Exploration and Center for Materials of the Universe, Arizona State University, Tempe, Arizona 85287, United States; Center for Materials of the Universe, Arizona State University, Tempe, Arizona 85281, United States

Tamilarasan Subramani − Center for Materials of the Universe and School of Molecular Sciences, Arizona State University, Tempe, Arizona 85281, United States; orcid.org/0000-0002-8121-1971

Kristina Lilova − Center for Materials of the Universe and School of Molecular Sciences, Arizona State University, Tempe, Arizona 85281, United States

James R. Lyons − Planetary Science Institute, Tucson, Arizona 85719, United States

Rhonda M. Stroud − School of Earth and Space Exploration and Center for Materials of the Universe, Arizona State University, Tempe, Arizona 85287, United States; Buseck Center for Meteorite Studies of Earth and Space Exploration, Tempe, Arizona 85827, United States

Complete contact information is available at:
https://pubs.acs.org/10.1021/acs.jpcc.3c04211

**Funding**

This work was supported by NASA grant 80NSSC22K1640.

**Notes**

The authors declare no competing financial interest.

## ACKNOWLEDGMENTS

We thank ASU faculty Hongwu Xu for help with GSAS-II analysis and Richard Hervig for valuable discussions and providing bulk OEn. We thank Karl Weiss, Kenneth Mossman, and Manuel Roldan-Gutierrez at the John M. Cowley Center for High Resolution Electron Microscopy (CHREM) at ASU for all electron microscopy work. We thank David Wright at the Goldwater Materials Science Facility (GMSF) at ASU for help with high-temperature synthesis. We acknowledge one reviewer for corroborating our pairwise values using a different calculation method.

## ABBREVIATIONS

OEn, orthoenstatite; PEn, protoenstatite; HCEn, high-clinoenstatite; LCEn, low-clinoenstatite; TEOS, tetraethyl orthosilicate; TG−DSC, thermogravimetry and differential scanning calorimetry; PXRD, powder X-ray diffraction; FTIR, Fourier transform infrared spectroscopy; ATR, attenuated reflectance spectroscopy; BET, Brunauer−Emmett−Teller; STEM, scanning transmission electron microscopy; TEM, transmission electron microscopy; SEM, scanning electron spectroscopy; FIB, focused ion beam; IA, interfacial area; SA, surface area; IE, interfacial energy; SE, surface energy; EM, electron microscopy; CAI, calcium−aluminum-rich inclusions; JWST, James Webb Space Telescope